\documentclass[prl,twocolumn,aps,preprintnumbers]{revtex4}

\usepackage{graphicx}

\begin{document}
\preprint{LBNL-49434}
{\bf Comment on ``Entry Distribution, Fission Barrier, and Formation
Mechanism of $ ^{254}_{102}$No ''}

\vspace{0.15cm}

A recent paper \cite{paper} has reported the observation of the
rotational band of 
$^{254}$No for spins up to $I$=20, showing that
the compound nucleus was formed and survived fission decay at angular
momenta $I\ge 20$. A microscopic description of $^{254}$No with the 
Gogny force predicts the observed survival of this nucleus against 
fission \cite{Egi00}.

  This finding may appear surprising at first, given
the well known instability toward fission of these nuclei
and the expected decrease in the fission barrier due to angular
momentum. The question behind the surprise 
is why angular momentum, usually so effective in decreasing the
fission barrier and in enhancing the fission to neutron emission
branching ratio in lighter nuclei, appears here to be somewhat
ineffective. The explanation of this puzzle is not only interesting
for this case, but is also even more relevant for the resilience to
angular momentum of superheavy nuclei.

To show the leading
effects of angular momentum on the barrier height, we use 
perturbation theory to calculate the energy associated with the 
perturbation (rotational energy) using parameter values (moments of 
inertia, deformation, etc.) associated with no perturbation. This is 
the standard cranking approximation. 
Accordingly, the energy can be written as: 
\begin{equation}
E\left(\vec\epsilon\right)=E_0\left(\vec\epsilon\right)+
\frac{I(I+1)\hbar^2}{2{\cal J}(\vec\epsilon
)}
\end{equation}
where $\vec\epsilon$ is a generalized deformation vector, 
$E_0(\vec\epsilon
)$ and ${\cal J}(\vec\epsilon )$ are the potential energy surface and 
moment of
inertia at zero angular momentum.

\begin{figure}
\includegraphics[width=3.2in]{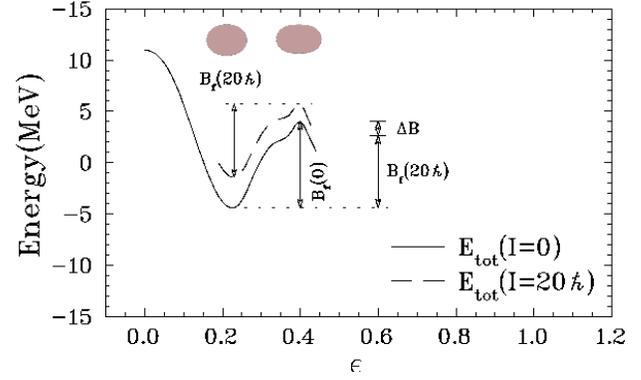}
\caption{A schematic description of the fission barrier for $^{254}$No. 
The solid line represents the total energy as a function of
deformation when $I=0$, while the dashed line is calculated for the 
value of
$I$ indicated in the figure. The shapes represent the ground state 
and saddle
configurations.}  \label{fig:fig1}
\end{figure}

As shown in Fig.~\ref{fig:fig1}, the decrease of the barrier height
$\Delta B(I)=B_f(I)-B_f(0)$ due to angular momentum is
\begin{equation}
\Delta B=\frac{\hbar ^2}{2}\left(\frac{1}{{\cal 
J}_{g}}-\frac{1}{{\cal J}_{s}}\right)I(I+1)
\end{equation}
where ${\cal J}_g$ and ${\cal J}_s$ are the moments of inertia of the 
ground and saddle deformations.
This decrease depends strictly on the values of the two moments of
inertia at $I$=0 irrespective of their origin (liquid drop, shell
effects, pairing, etc.). Higher order effects, such as changes in the
ground and saddle deformation, and changes in the shell and pairing effects 
occur
at higher angular momenta. The evidence for the goodness of the 
cranking approximation and thus for the lack of change of the shell 
effect in the ground state is evident in the fact that $^{254}$No (as 
well as most strongly deformed rare earth and actinide nuclei) is a 
good rigid rotor up to $I$=20. The moment of inertia changes by 
$\approx$10\% for $I$=0 to $I$=20.

In typical lighter nuclei, the saddle point is controlled by the
liquid drop contributions and is found to be at large
deformations. Therefore,
${\cal J}_{g}<<{\cal J}_{s}$
and
$\Delta B\approx\hbar^2I(I+1)/{2{\cal J}_{g}}$.
This produces the large effect of angular momentum on the fission
barrier for lighter systems. 

However, in trans-Fermium nuclei, the ground state
is already deformed at the values of $\epsilon$ typical of all
actinides. The saddle occurs at a deformation only slightly greater,
corresponding to the anti-shell immediately following the deformed 
minimum.
We can rewrite Eq.~(2) in terms of the fractional difference of the 
moments of inertia:
small values of $\Delta {\cal J}={\cal J}_s-{\cal J}_g$, obtaining
\begin{equation}
\Delta B=\frac{\hbar^2I(I+1)}{2{\cal J}_{g}}\frac{\Delta {\cal 
J}}{{\cal J}_{s}}=E_{rot}^{gs}\frac{\Delta{\cal J}}{{\cal J}_{s} }
\end{equation}
where $\Delta {\cal J}={\cal J}_s-{\cal J}_g$.  Consequently, the
decrease in barrier height is equal to the ground state rotational
energy ($E_{rot}^{gs}$) times the fractional change in the moment of
inertia. For $I$=20, the rotational energy $E_{rot}^{gs}\approx$2.8
MeV and $\Delta {\cal J}/{\cal J}_s\approx 0.40$ giving $\Delta B$
$\approx 1.1$ MeV. These features are shown pictorially in Fig.~1.
This estimate (1.1 MeV) of the change in fission barrier at 20$\hbar$
is in excellent agreement with detailed calculations \cite{Egi00}.

There is little doubt that the estimate described above explains the
resiliency of the No barrier to angular momentum. The same arguments
speak for a similar or even greater resilience in superheavy
nuclei. Thus, we expect that we can safely graze in the pastures of
the superheavy island of stability, without fear of (moderate) angular
momentum values.

\vspace{0.2cm}
\noindent
L.G. Moretto, L. Phair, and G.J. Wozniak


Lawrence Berkeley National Laboratory

Berkeley, California 94720



\vspace{-0.6cm}
\begin{thebibliography}{99}
\bibitem{paper} P. Reiter {\em et al.}, Phys. Rev. Lett. {\bf 84}, 
3542 (2000).
\bibitem{Egi00} J.L. Egido and L.M. Robledo, Phys. Rev. Lett. {\bf 
85}, 1198 (2000).

\end{thebibliography}
\end{document}